\documentclass[abstract=on]{scrreprt}

\author{Justus Fasse\\justus.fasse@kuleuven.be \and Bart Jacobs\\bart.jacobs@kuleuven.be}
\publishers{imec-DistriNet, KU Leuven, 3001 Leuven, Belgium}

\title{Modular termination verification with a higher-order concurrent separation logic}
\subtitle{\textsc{Intermediate report}
\renewcommand\footnotemark{} % Do not use asterisk for funding notice
\thanks{This research is partially funded by the Research Fund KU Leuven, and by the Flemish Research Programme Cybersecurity.}}

\usepackage{biblatex}
\usepackage[T1]{fontenc}
\usepackage[utf8]{inputenc}
\usepackage{FiraSans}

\usepackage{amsmath}

\usepackage{graphicx}
\graphicspath{{pictures/}}

\usepackage{amsthm}
\newtheorem{defn}{Definition}

\newtheorem{lem}{Lemma}
\newtheorem{thm}{Theorem}

\usepackage{amssymb}
\usepackage{mathpartir}

\usepackage{cancel}

\usepackage{xspace}

\newcommand{\heaplang}{HeapLang\xspace}
\newcommand{\coq}{Coq\xspace}
\newcommand{\heaplanglt}{HeapLang\textsuperscript{<}\xspace}
\newcommand{\burn}{\langkw{Burn}\xspace}
\newcommand{\burns}{\burn{}s\xspace}
\newcommand{\pseudosize}{\mathit{pseudo\_size}}
\newcommand{\nbunprotectedapps}{\mathit{nb\_unprotected\_apps}}
\newcommand{\enoughburnsexpr}{\mathit{enough\_burns\_expr}}
\newcommand{\enoughburnsval}{\mathit{enough\_burns\_val}}
\newcommand{\enoughburnsheap}{\mathit{enough\_burns\_heap}}
\newcommand{\enoughburnscfg}{\mathit{enough\_burns\_cfg}}

\usepackage{listings}
\usepackage{iris-latex/heaplang}
\usepackage{iris-latex/iris}

\usepackage{hyperref}
\begin{document}

\maketitle

\abstract{
We report on intermediate results of our research on reasoning about liveness properties in addition to deep correctness properties for an imperative, concurrent programming language with a higher-order store.
At present, we focus on one particular liveness property, namely termination.
By guaranteeing termination we can strengthen statements of partial correctness to total correctness.
This is achieved by the classic approach of turning termination into a safety property.
In particular we extend the programming language under consideration with call permissions, which have been shown to enable modular reasoning about termination.
Atomic blocks are added to increase the expressiveness of our call-permission-based approach.
Our work builds on top of Iris---a foundational, machine-checked, higher-order concurrent separation logic framework---without modifying it.
With these additions we are able to modularly reason about the termination of concurrent, but non-blocking algorithms.
Our additions to the programming language under consideration preserve Iris' ability to reason about helping and prophecies.
As an example, we apply the current system to an existing case study for a lock-free concurrent stack with helping that has been proven in Iris.
Finally, we sketch the next steps to scale our approach to blocking concurrency.
}

\tableofcontents

\chapter{Introduction}

Most work on Hoare-style~\autocite{DBLP:journals/cacm/Hoare69} program verification (i.e.\ verification based on pre- and postconditions) so far has dealt only with partial correctness, that is proving statements of the form: if the program is run in a state satisfying the precondition the program is safe, and---\emph{if it terminates}---the postcondition will hold.
The field of program verification has made great strides in verifying ever more complex programs.
Separation logic~\autocite{DBLP:conf/lics/Reynolds02}, and in particular concurrent separation logic~\autocite{DBLP:journals/tcs/OHearn07} have been shown to scale to concurrent programs in a modular way.
Recent work tackling increasingly daring programming language features and patterns has culminated in the Iris framework~\autocite{iris,iris2,DBLP:conf/esop/Krebbers0BJDB17,DBLP:journals/pacmpl/SpiesGTJKBD22}.
Unfortunately, however, Iris too does not support total correctness reasoning.
With few exceptions---some of which will be discussed in Section~\ref{sec:related-work}---all Iris-based projects include the caveat that they have to assume termination.
In this project, we aim to enable modular total correctness reasoning in Iris.
That is, instead of having to assume termination, we wish to prove it thus strengthening the guarantees given by our Hoare triples.
This preliminary report presents our approach for reasoning about total correctness of non-blocking programs. It also sketches our ideas for extending this result to blocking programs.

We are working with Iris 4.0~\autocite{iris,iris2,DBLP:conf/esop/Krebbers0BJDB17,DBLP:journals/pacmpl/SpiesGTJKBD22}, a mechanized, foundational, higher-order concurrent separation logic.
Working with such a logic (framework) is interesting for many reasons:
\begin{itemize}
  \item
    Mechanized proofs provide the highest level of trust in our proofs because their correctness is automatically rigorously checked on each run.
    Moreover, they can drastically aid the social process to accept proofs~\autocite{DBLP:journals/cacm/DeMilloLP79} by reducing the trusted computing base to a small kernel~\autocite{DBLP:journals/cacm/Vardi21c}. 
    Proof assistants, among them \coq, usually rely on a particularly small trusted computing base.
    See \autocite[\S 2.1]{DBLP:conf/esop/MonniauxB22} for a short discussion on the reliability of Coq's proof checking.
  \item In this setting, foundational refers to the property of verifying directly ``against the operational semantics of the programming languages under
  consideration and assuming only the low-level axioms of mathematical logic (as encoded in the type theory of Coq).''~\autocite{iris-ground-up}
  \item Concurrent separation logic allows us to modularly reason about concurrent, imperative programs with its concepts of (flexible) ownership and invariants (cf.\ \autocite{iris}).
  \item Higher-order logic and impredicative invariants support important use cases:
    \begin{itemize}
      \item Languages with dynamic allocation and a higher-order store (general references)~\autocite{DBLP:conf/popl/AhmedDR09,iris}
      \item General recursive types~\autocite{appel-mcallester}
      \item Higher-order modular specifications~\autocite{hocap} and the ability to encode logical atomicity~\autocite{iris}
      \item Nested Hoare-triples to, for example, reason about returned closures~\autocite{iris-ground-up} and higher-order functions~\autocite{iris-lecture-notes}
    \end{itemize}
\end{itemize}
Please refer to \citetitle{iris-ground-up}~\autocite{iris-ground-up} for more details on the history and motivations behind Iris.
We further note that by using Iris we inherit its support for interactive theorem proving~\autocite{DBLP:conf/popl/KrebbersTB17}.

In order to deal with the semantic circularity that the aforementioned higher-order features introduce, Iris employs step indexing~\autocite{appel-mcallester,ahmed-thesis,birkedal+:lics11,DBLP:journals/corr/abs-1103-0510}.
Step-indexed logics give semantics to programs based on arbitrarily long but finite prefixes of their behavior and are thus a priori limited to safety~\autocite{DBLP:journals/ipl/AlpernS85} properties~\autocite{DBLP:conf/pldi/SpiesGGTKDB21}.
As a result, a huge class of properties, namely liveness properties, fall outside the purview of Iris.
This includes, but is not limited to, properties of the form ``eventually \ldots'', e.g.\ the program will eventually terminate; and ``response'' properties, e.g.\ the server will always eventually respond to a client's request.
More generally, the field of temporal logic(s) (starting with~\autocite{DBLP:conf/focs/Pnueli77}) is dedicated to this important class of properties.
We seek to address this shortcoming---for the time being focusing on termination---in such a way as to provide modular (Hoare-style) specifications that ensure the safe termination of programs.

Thus, this report presents intermediate results towards the foundational, mechanized verification of termination for concurrent, imperative programs.

\section{High-level overview}%
\label{sec:overview}

We define an instrumented semantics and a program instrumentation scheme (``ghost code'') such that if an instrumented version of the program does not get stuck in the instrumented semantics, the original program terminates in the original semantics.
We instantiate Iris with the instrumented semantics to prove the former, and then apply our meta-theorem\footnote{Not yet formalized.} to obtain the latter.

We now present this idea in more detail by considering a language whose only source of non-termination is function application/recursion (but including e.g.\ Landin's knot, i.e.\ tying the recursive knot via the higher-order store).
This is not a stretch because we can reformulate loops as recursive functions.
In such a setting, we can prove termination if we can ensure that the program will only perform a finite number of function applications.
We take inspiration from \citeauthor{DBLP:journals/toplas/JacobsBK18}~\autocite{DBLP:journals/toplas/JacobsBK18} and transport the idea to the setting of a higher-order separation logic, in particular Iris~\autocite{iris}.
The basic setup of \autocite{DBLP:journals/toplas/JacobsBK18} looks as follows: we consider for the programming language in question an instrumented semantics that ``burns'' at every function application an angelically picked call permission from a finite stock (multiset) of call permissions.
In the instrumented semantics, if the multiset of call permissions is empty when a function call is attempted, the program gets stuck.
To facilitate modular specification, call permissions are qualified by a \emph{level}, an element of some well-founded set of levels.
The instrumented semantics may at any time angelically choose to replace a call permission at some level $L$ by some finite number of call permissions at some level $L' < L$.
By choosing the names of the program's modules (functions/procedures/classes/\ldots) as levels, ordered by their order in the program text (or the module dependency graph), and requiring in a module's precondition a call permission qualified by that module's name, a module can always perform arbitrarily many calls into earlier modules (``upcalls'').
In the following the terms ``call permission at level $L$'' and ``call permission $L$'' will be used interchangeably.
The replacement scheme described above conforms to the well-founded multiset order as defined by \citeauthor{DBLP:journals/cacm/DershowitzM79}~\autocite{DBLP:journals/cacm/DershowitzM79}.
It follows that, if an instrumented execution of the program does not go wrong, then the execution's erasure---recovering the original semantics---terminates.
Thus, proving for every original execution the existence of a safe instrumented execution implies termination of the program under consideration in the original semantics.

Unlike \citeauthor{DBLP:journals/toplas/JacobsBK18}~\autocite{DBLP:journals/toplas/JacobsBK18}, we manifest the stock of call permissions in an extended version of the programming language we are ultimately targeting.
Furthermore, instead of having angelic nondeterminism in the semantics, we extend the syntax of the programming language with a command (called \burn) for explicitly replacing a call permission at a specified level $L$ by a specified number of call permissions at a specified lower level $L'$.
This decision is related to the issues of (finite) step-indexing in Iris as discussed by e.g.\ \citeauthor{DBLP:conf/pldi/SpiesGGTKDB21}~\autocite{DBLP:conf/pldi/SpiesGGTKDB21}.
In short: if we were to allow (angelic) control over how to lower call permissions at the program-logic level, we could ``prove'' the termination of non-terminating programs rendering our approach unsound and effectively useless.
Consider a non-terminating function, e.g.\ $\omega = (\lambda x. x~x), \Omega = \omega~\omega$ and the presence of an arbitrary call permissions $cp$ that is larger (qualified by a higher level) than another one, $cp'$.
Essentially, we could remove $cp$ from the multiset of call permissions and create $n + 1$ instances of call permission $cp'$, where $n$ is the (arbitrary but finite) step-index/prefix of the computation.
Now, we could show that $\Omega$ does not get stuck because we can justify $n + 1$ function applications and incorrectly conclude that it ``terminates''.
By eliminating angelic nondeterminism, we guarantee that our choice cannot possibly depend on the step-index.

Basically we decompose the proof of termination into two (three) parts.
\begin{enumerate}
  \item [(0.)]
    We require the programmer to instrument the original program with \burns in such a way that every infinite execution would entail the execution of infinitely many \burn instructions; an impossibility (see Section~\ref{sec:heaplanglt-termination}). More precisely we require the program to pass the verified checking function described next.
  \item
    We define a verified-correct (but conservative) checking function that, when returning true, guarantees that given the inserted \burns the instrumented program will either successfully reduce to a value or get stuck.
    This is a meta-theoretical result about the language we have defined, independent of Iris and its step indexing.
  \item
    We use Iris' partial correctness logic to prove safety (as well as deep correctness properties that can be expressed as safety properties) of our program.
    Finally---but missing from the current \coq development---we need an erasure that strips the \burns from the program, together with a corresponding proof that the properties of functional correctness and termination are transported.
\end{enumerate}
The safety proof of part two, strengthens the statement of termination of part one to \emph{safe} termination.
Similarly, the second result's partial correctness properties are turned into total correctness properties when used in conjunction with part one.

We will use square bracket to denote multisets.
For example, $[1] \neq [1,1]$.
Moreover, by abuse of notation, we will write the multiset of $n$ ones as follows: $[n \cdot 1]$.
The mentioned proofs as well as all intermediate lemmas are formalized and proven with \coq and can be found in the code accompanying this report\footnote{\url{https://doi.org/10.5281/zenodo.7488894}}.

Section~\ref{sec:heaplanglt} discusses the additions to Iris' default language \heaplang resulting in our programming language \heaplanglt; the ensuant changes in the program logic are discussed in Section~\ref{sec:heaplanglt-program-logic}.
These two sections lay the groundwork for the proof sketch for a concurrent stack with helping (Section~\ref{sec:concurrent-stack-with-helping}), which we extend to prove termination of the operations on it.
Finally, Section~\ref{sec:related-work} briefly discusses some related work and Section~\ref{sec:outlook} concludes with an outlook on the work left to do.

\chapter{\heaplanglt}%
\label{sec:heaplanglt}

\begin{figure}
\begin{align*}
  \loc \ni \Loc \eqdef{}& \integer \\
  \prophid \ni \ProphId \eqdef{}& \integer \\
  \textcolor{blue}{cp \ni \textdom{CP} \eqdef{}}& \textcolor{blue}{\nat} \\
  \sigma \ni \State \eqdef{}& \record{\begin{aligned}
                   \stateHeap:{}& \Loc \fpfn \Val \\
                   \stateProphs:{}& \pset{\ProphId} \\
		   \textcolor{blue}{\textsc{CallPerms}}:{}& \textcolor{blue}{\textdom{Multiset}(\textdom{CP})}
                 \end{aligned}} \\
  % \obs \ni \Obs \eqdef{}& \ProphId \times (\Val \times \Val)
\end{align*}
\caption{Adapted state for \heaplanglt programs.
Our extension to the state definition is highlighted in blue.
Note that currently the type of individual call permissions, \textdom{CP}, is defined as \( \nat \).
However, any type with a well-founded order on its elements can be used.}%
\label{fig:state-def}
\end{figure}

\heaplanglt is a conservative extension of the default language shipped with Iris, called \heaplang~\autocite{iris-technical-reference-4-0}.
It is conservative in the sense that every \heaplang program is also a \heaplanglt program and has the same semantics.
They are ``ML-like languages with a higher-order heap, unstructured concurrency, some common atomic operations, and prophecy variables.''~\autocite{iris-technical-reference-4-0}
The mentioned prophecy variables and related ghost code were added to \heaplang ``to verify \emph{logical atomicity} (a relative of linearizability) for classic examples from the concurrency literature like RDCSS and the Herlihy-Wing queue''~\autocite{DBLP:journals/pacmpl/JungLPRTDJ20}.
More details on \heaplang can be found in Chapter 12 of Iris' technical reference~\autocite{iris-technical-reference-4-0}.
\heaplanglt extends \heaplang with two orthogonal but complementary features: call permissions and atomic blocks.
The necessity of adding atomic blocks for our approach is explained in Section~\ref{sec:concurrent-stack-with-helping} with the example of a concurrent stack with helping.
Additionally, we show that they are more generally useful.
For example, basic atomic primitives such as compare-and-swap (CAS) can be encoded with them.
We proved adequacy of Iris for \heaplanglt; the theorem and proof are straightforward adaptations from \heaplang.

Using the notion of call permissions and how to reduce them (\langkw{Burn}) (see Section~\ref{sec:heaplanglt-call-permissions}) we define our instrumentation scheme to guarantee termination in Section~\ref{sec:heaplanglt-termination}.
A \heaplanglt program that is safe with regard to Iris' program logic that also passes the check of Section~\ref{sec:heaplanglt-termination} is guaranteed to eventually reduce to a value (safely terminate).
The erasure of such a program produces the same value without the notion of call permissions, which are tracked separately from the other program state (see Figure~\ref{fig:state-def}) and are not acted upon by \heaplang operations.
If an instrumented \heaplanglt program terminates safely, i.e.\ reduces to a value, its erased version---a \heaplang program\footnote{Modulo the atomic block construct we introduce.
Interestingly, we can consider an atomic block consisting of a single atomic instruction as no-op. Such a situation occurs in our example (Section~\ref{sec:concurrent-stack-with-helping}) after erasure.}---will also terminate safely to the same value.
We therefore prove termination outside of Iris and transfer this property via erasure\footnote{The erasure and related proofs is not yet implemented in the \coq development.} to the desired program.

\section{Call permissions}%
\label{sec:heaplanglt-call-permissions}

The extended \heaplanglt state definition is shown in Figure~\ref{fig:state-def}.
A single call-permissions-manipulating instruction is added to the syntax and semantics of \heaplanglt: \burn.
All other instructions of \heaplanglt are the same as in \heaplang and are not affected by the extension of the state definition.
\( \langkw{Burn}~\expr_1~cp~\expr_2~cp' \) consumes a call permission $cp$ and produces the (non-negative) number of the lower call permission $cp'$ computed by $\expr_2$.
It then reduces to the wrapped expression $e_1$.
The following notation is used for this replacement: ${\state : \textsc{CallPerms}[cp \rightarrow n \cdot cp']}$ describes the state $\state'$ such that ${\state'.\stateHeap = \state.\stateHeap}$ and ${\state'.\stateProphs = \state.\stateProphs}$ and ${\state'.\textsc{CallPerms} = \state.\textsc{CallPerms} \setminus [cp] \uplus [n \cdot cp']}$.
The arbitrary but finite number is itself determined by a \heaplanglt expression thus precluding the inspection of the step-index in the logic but remaining pretty flexible.
The expression will get stuck if
\begin{itemize}
  \item $\expr_2$ gets stuck or evaluates to anything other than a non-negative integer or
  \item $cp' \nless cp$ or
  \item $cp \notin \sigma.\textsc{CallPerms}$
\end{itemize}

\begin{figure}
\begin{align*}
\val,\valB \in \Val \bnfdef{}&
  z \mid
  \True \mid \False \mid
  \TT \mid
  \poison \mid
  \loc \mid
  \prophid \mid {}& (z \in \integer, \loc \in \Loc, \prophid \in \ProphId) \\&
  \RecV\lvarF(\lvar)= \expr \mid
  (\val,\valB)_\valForm \mid
  \Inl(\val)_\valForm \mid
  \Inr(\val)_\valForm  \\
\expr \in \Expr \bnfdef{}&
  \val \mid
  \lvar \mid
  \RecE\lvarF(\lvar)= \expr \mid
  \expr_1(\expr_2) \mid
  {}\\ &
  \HLOp_1 \expr \mid
  \expr_1 \HLOp_2 \expr_2 \mid
  \If \expr then \expr_1 \Else \expr_2 \mid
  {}\\ &
  (\expr_1,\expr_2)_\exprForm \mid
  \Fst(\expr) \mid
  \Snd(\expr) \mid
  {}\\ &
  \Inl(\expr)_\exprForm \mid
  \Inr(\expr)_\exprForm \mid
  \Match \expr with \Inl => \expr_1 | \Inr => \expr_2 end \mid
  {}\\ &
  \Alloc(\expr_1,\expr_2) \mid
  \Free(\expr) \mid
  \deref \expr \mid
  \expr_1 \gets \expr_2 \mid
  \CmpXchg(\expr_1, \expr_2, \expr_3) \mid
  \Xchg(\expr_1, \expr_2) \mid
  \FAA(\expr_1, \expr_2) \mid
  \kern-30ex{}\\ &
  \Fork \expr \mid
  \NewProph \mid
  \ResolveWith \expr_1 at \expr_2 to \expr_3 \;\textcolor{blue}{\mid} \\ &
  \textcolor{blue}{\langkw{Burn}(\expr_1, cp, \expr_2, cp')} \mid \textcolor{blue}{\langkw{AtomicBlock}(\expr)} & (\textcolor{blue}{cp, cp' \in \textdom{CP}}) \\
\HLOp_1 \bnfdef{}& - \mid \ldots ~~\text{(list incomplete)} \\
\HLOp_2 \bnfdef{}& + \mid - \mid \Ptradd \mid \mathop{=} \mid \ldots ~~\text{(list incomplete)}
\end{align*}
% (Note that \langkw{match} contains a literal $|$ that is not part of the BNF but part of HeapLang syntax.)
\caption{The modified syntax of \heaplanglt, additions are shown in blue.}%
\label{fig:heaplanglt-syntax}
\end{figure}

\begin{figure}
\judgment[``Head'' reduction (impure)]{\expr_1, \state_1 \hstep [\vec\obs] \expr_2, \state_2, \vec\expr}
\newcommand\alignheader{\kern-30ex}
\begin{align*}
&\alignheader\textbf{Heap reductions} \\
(\Alloc(z, \val), \state) \hstep[\nil]{}&
  (\loc, \mapinsert {[\loc,\loc+z)} \val {\state:\stateHeap}, \nil)
  &&\text{if $z>0$ and} \\
  &&& \text{\(\All i<z. \state.\stateHeap(\loc+i) = \bot\)} \\
(\Free(\loc), \state) \hstep[\nil]{}&
  ((), \mapinsert \loc \bot {\state:\stateHeap}, \nil) &&\text{if $\state.\stateHeap(l) = \val$} \\
(\deref\loc, \state) \hstep[\nil]{}&
  (\val, \state, \nil) &&\text{if $\state.\stateHeap(l) = \val$} \\
(\loc\gets\valB, \state) \hstep[\nil]{}&
  (\TT, \mapinsert \loc \valB {\state:\stateHeap}, \nil)  &&\text{if $\state.\stateHeap(l) = \val$} \\
(\CmpXchg(\loc,\valB_1,\valB_2), \state) \hstep[\nil]{}&
  ((\val, \True), \mapinsert\loc{\valB_2}{\state:\stateHeap}, \nil)
  &&\text{if $\state.\stateHeap(l) = \val$ and $\val \valeq \valB_1$} \\
(\CmpXchg(\loc,\valB_1,\valB_2), \state) \hstep[\nil]{}&
  ((\val, \False), \state, \nil)
  &&\text{if $\state.\stateHeap(l) = \val$ and $\val \valne \valB_1$} \\
(\Xchg(\loc, \valB) \hstep[\nil]{}&
  (\val, \mapinsert \loc \valB {\state:\stateHeap}, \nil) &&\text{if $\state.\stateHeap(l) = \val$} \\
(\FAA(\loc, z_2) \hstep[\nil]{}&
  (z_1, \mapinsert \loc {z_1+z_2} {\state:\stateHeap}, \nil) &&\text{if $\state.\stateHeap(l) = z_1$} \\
&\alignheader\textbf{Call permissions manipulation} \\
\textcolor{blue}{(\langkw{Burn}(\expr,cp,n,cp'), \state) \hstep[\nil]}{}&
  \textcolor{blue}{(\expr, {\state : \textsc{CallPerms}[cp \rightarrow [n \cdot cp']]}, \nil)}
  &&\text{\textcolor{blue}{if $cp \in \state.\textsc{CallPerms}$ and}} \\
  &&&\text{\textcolor{blue}{$cp' < cp$ and $0 \leq n$}} \\
&\alignheader\textbf{Special reductions} \\
\textcolor{blue}{(\langkw{AtomicBlock}(\expr), \state) \hstep[\nil]}{}&
  \textcolor{blue}{(v, \state',\nil)} &&\text{\textcolor{blue}{if $e,\state \Rightarrow v,\state'$}} \\
(\Fork\expr, \state) \hstep[\nil]{}&
  (\TT, \state, \expr) \\
(\NewProph, \state) \hstep[\nil]{}&
  (\prophid, \state:\stateProphs \uplus \set{\prophid}, \nil)
  &&\text{if $\prophid \notin \state.\stateProphs$}
\end{align*}
\begin{mathpar}
\infer
  {(\expr, \state) \hstep[\vec\obs] (\val, \state', \vec\expr')}
  {(\ResolveWith \expr at \prophid to \valB, \state) \hstep[\vec\obs \dplus [(\prophid, (\val, \valB))]] (\val, \state', \vec\expr')}
\end{mathpar}
\caption{The operational semantics of \heaplanglt with our added constructs shown in blue. \langkw{AtomicBlock} makes use of our big-step semantics for a subset of \heaplanglt (See Figure~\ref{fig:eval-big-step}).
Notice how the side condition on \langkw{Burn} ensures that the multiset of call permissions becomes strictly smaller.
Recall the notation for call permission state updates introduced in Section~\ref{sec:heaplanglt-call-permissions}.
}
\label{fig:heaplang-reduction-impure}
\end{figure}

\begin{figure}
\begin{mathpar}
  \inferrule
    { }
    {v_\exprForm,\state \Rightarrow v_\valForm,\state}
  \and \inferrule
    { }
    {\RecE\lvarF(\lvar)= \expr, \state \Rightarrow \RecV\lvarF(\lvar)= \expr, \state }
  \and \inferrule
    { \expr,\state \Rightarrow z, \state' }
    { (-_{\HLOp} e, \state) \Rightarrow -z, \state' }
  \and \inferrule
    { \expr_1,\state' \Rightarrow z_1, \state'' \\ \expr_2,\state \Rightarrow z_2, \state' }
    { \expr_1 +_{\HLOp} \expr_2, \state \Rightarrow z_1 + z_2, \state'' }
  \and \inferrule
    { \expr_1,\state' \Rightarrow z_1, \state'' \\ \expr_2,\state \Rightarrow z_2, \state' }
    { \expr_1 -_{\HLOp} \expr_2, \state \Rightarrow z_1 - z_2, \state'' }
  \and \inferrule
    { \expr_1,\state' \Rightarrow l, \state'' \\ \expr_2,\state \Rightarrow z, \state' }
    { \expr_1 \Ptradd \expr_2, \state \Rightarrow \loc + z, \state'' }
  \and \inferrule
    { \expr_1,\state' \Rightarrow \val_1, \state'' \\ \expr_2,\state \Rightarrow \val_2, \state' \\\\
      b = \text{if $\val_1 \valeq \val_2$ then $\True$ else if $\val_1 \valne \val_2$ then $\False$ } }
    { \expr_1 =_{\HLOp} \expr_2, \state \Rightarrow b, \state'' }
  \and \inferrule
    { \state'.\stateHeap(l) = \val \\ \expr,\state \Rightarrow l,\state'}
    { \deref\expr,\state \Rightarrow \val,\state' }
  \and \inferrule
    { \state.\stateHeap(l) = \val \\ \expr_1,\state' \Rightarrow \loc, \state'' \\ \expr_2,\state \Rightarrow \valB,\state' }
    { \expr_1\gets\expr_2,\state \Rightarrow \TT,\mapinsert \loc \valB {\state'':\stateHeap} }
  \and \inferrule
    { cp' < cp \\
      cp \in \state'.\textsc{CallPerms} \\
      \expr_2,\state \Rightarrow n,\state' \\\\
      \expr_1,{\state' : \textsc{CallPerms}[cp \rightarrow [n \cdot cp']]} \Rightarrow \val,\state'' \\ }
    { \langkw{Burn}(\expr_1,cp,\expr_2,cp'),\state \Rightarrow v,\state'' }
  \and \inferrule
    { \expr_1,\state' \Rightarrow \RecV\lvarF(\lvar)= \expr,\state'' \\ \expr_2,\state \Rightarrow \val,\state' \\\\
      \subst {\subst \expr \lvarF {(\Rec\lvarF(\lvar)= \expr)}} \lvar \val,\state'' \Rightarrow \valB,\state''' }
    { \expr_1~\expr_2,\state \Rightarrow \valB,\state''' }
\end{mathpar}
\caption{A big-step evaluation relation for a subset of \heaplanglt.
\emph{The premises are to-be-read right to left}, matching the evaluation order of \heaplanglt.
A corresponding relation does not exist for \heaplang and therefore the additions (the entire figure) are not shown in blue.
We separately show in the \coq development that this relation is deterministic.
Importantly (cf.\ Section~\ref{sec:concurrent-stack-with-helping}), evaluations of \burns is included which allows us to perform them simultaneously with other operations.
Recall the notation for a call permissions update (Figure~\ref{fig:heaplang-reduction-impure}).}%
\label{fig:eval-big-step}
\end{figure}

\begin{figure}
\begin{align*}
\lctx \in \Lctx \bnfdef{}&
  \bullet \mid \Lctx_{>} \\
\lctx_{>} \in \Lctx_{>} \bnfdef{}&
  \expr(\lctx) \mid
  \lctx (\val) \mid
  {}\\ &
  \HLOp_1 \lctx \mid
  \expr \HLOp_2 \lctx \mid
  \lctx \HLOp_2 \val \mid
  \If \lctx then \expr_1 \Else \expr_2 \mid
  {}\\ &
  (\expr, \lctx) \mid
  (\lctx, \val) \mid
  \Fst(\lctx) \mid
  \Snd(\lctx) \mid
  {}\\ &
  \Inl(\lctx) \mid
  \Inr(\lctx) \mid
  \Match \lctx with \Inl => \expr_1 | \Inr => \expr_2 end \mid
  {}\\ &
  \Alloc(\expr, \lctx) \mid
  \Alloc(\lctx, \val) \mid
  \Free(\lctx) \mid
  \deref \lctx \mid
  \expr \gets \lctx \mid
  \lctx \gets \val \mid
  {}\\ &
  \CmpXchg(\expr_1, \expr_2, \lctx) \mid
  \CmpXchg(\expr_1, \lctx, \val_3) \mid
  \CmpXchg(\lctx, \val_2, \val_3) \mid
  {}\\ &
  \Xchg(\expr, \lctx) \mid
  \Xchg(\lctx, \val) \mid
  \FAA(\expr, \lctx) \mid
  \FAA(\lctx, \val) \mid
  {}\\ &
  \ResolveWith \expr_1 at \expr_2 to \lctx \mid
  \ResolveWith \expr_1 at \lctx to \val_3 \mid
  {}\\ &
  \ResolveWith \lctx_{>} at \val_2 to \val_3 \\ &
  \textcolor{blue}{\langkw{Burn}(\expr, cp,\lctx, cp')} \mid \textcolor{blue}{\langkw{AtomicBlock}(\expr)}
\end{align*}
\caption{Adapted (additions in blue) evaluation contexts. Note that the expression computing the number of times $cp'$ should be created is an evaluation context while the expressions wrapped by \burn and \langkw{AtomicBlock} are not.}%
\label{fig:eval-contexts}
\end{figure}

The modified (abstract) syntax for \heaplanglt that includes the \burn construct is shown in Figure~\ref{fig:heaplanglt-syntax}.
\heaplang and therefore \heaplanglt follow a right-to-left evaluation strategy, defined via evaluation contexts (see Figure~\ref{fig:eval-contexts}).
The expression protected by the \burn is naturally not evaluated before the \burn itself, while the expression determining the number of lowered call permissions to insert must be.
Finally, the head-step semantics of \burn is given in Figure~\ref{fig:heaplang-reduction-impure}, matching the aforementioned description.

The remaining---unmodified---rules, in particular for ``pure'' (non-state manipulating) steps and the thread-interleaving semantics remain unchanged and are omitted from this report.
The definition of program configuration will be used in the discussion of our instrumentation scheme in Section~\ref{sec:heaplanglt-termination}:
\begin{defn}
  A program configuration $\rho$  of type \textdom{cfg} is a pair of a) a list of expressions ($\rho.\textsc{threads}$), each expression representing a thread and b) the shared global state ($\rho.\textsc{state}$, see Figure~\ref{fig:state-def}).
\end{defn}

\section{Atomic blocks}%
\label{sec:heaplanglt-atomic-blocks}

Atomic blocks define, as the name suggests, blocks of code that will be executed atomically, that is in a single step.
More formally, an atomic block wraps an expression, and returns that expression's resulting value, as well as applying all state changes, in a single step.
The wrapped expression must not get stuck, and result in a value.
Otherwise the atomic block gets stuck itself.
Our atomic blocks are limited in the expressions they support.
For example we disallow the forking of threads and heap allocations.
The supported expressions are defined via the big-step evaluation relation shown in Figure~\ref{fig:eval-big-step} that is used to evaluate the atomic block.

The syntax and semantics of atomic blocks is given in Figures~\ref{fig:heaplanglt-syntax} (syntax), \ref{fig:eval-contexts} (evaluation order) and~\ref{fig:heaplang-reduction-impure} (impure head-step relation).
They are adaptions from the Iris 4.0 Reference~\autocite{iris-technical-reference-4-0} section about \heaplang.

\section{Instrumentation scheme}%
\label{sec:heaplanglt-termination}

\begin{figure}
\begin{align*}
  \pseudosize~:~\textdom{expr} &\rightarrow \nat \\
  \pseudosize~(v : val) &= 0 \\
  \pseudosize~(x : var) &= 1 \\
  \pseudosize~(\langkw{AtomicBlock}~\_) &= 1 \\
  \pseudosize~\NewProph &= 1 \\
  \pseudosize~(\RecE\lvarF(\lvar)= \expr) &= 1 + \pseudosize~\expr \\
  \pseudosize~(\Match \expr with \Inl => \expr_1 | \Inr => \expr_2 end) &= 2 + \pseudosize~\expr + \\
  &\qquad \pseudosize~\expr_1 + \pseudosize~\expr_2 \\
  \pseudosize~e &= 1 + \sum_{i=1}^{n} \pseudosize~\expr_i \\
  & \qquad \text{where $\expr_1 \ldots \expr_n$ are the sub-expressions of $e$}
\end{align*}
  \caption{Function calculating our pseudo size for expressions. The pseudo size of $\langkw{match}$ adds two because the match is reduced to a function application, not a strict sub-expression.}%
\label{fig:pseudo-size-def}
\end{figure}

\begin{figure}
\begin{align*}
  \nbunprotectedapps~&:~\textdom{expr} \rightarrow \nat \\
  \nbunprotectedapps~(\RecE~\_(\_)=~\_) &= 0 \\
  \nbunprotectedapps~(\langkw{Burn}(\expr_1, cp, \expr_2, cp') &= \nbunprotectedapps~\expr_2 \\
  \nbunprotectedapps~(\expr_1(\expr_2)) &= 1 + \nbunprotectedapps~\expr_1~+ \\
                                           &\qquad \nbunprotectedapps~\expr_2 \\
  \nbunprotectedapps~(v : val) &= 0 \\
  \nbunprotectedapps~(x : var) &= 0 \\
  \nbunprotectedapps~\NewProph &= 0 \\
  \nbunprotectedapps~e &= \sum_{i=1}^{n} \nbunprotectedapps~\expr_i \\
  & \qquad \text{where $\expr_1 \ldots \expr_n$ are the sub-expressions of $e$}
\end{align*}
\caption{Overapproximating the number of unprotected function applications.
  Here, recursion into the body of recursive function definitions is skipped.
  Its use in Figure~\ref{fig:enough-burns-def} ensures that all function definitions contain enough \burns.}%
\label{fig:nb-unprotected-apps-def}
\end{figure}

\begin{figure}
\begin{align*}
  \enoughburnsexpr~:~\textdom{expr} \rightarrow \bool~&\text{and}~\enoughburnsval~:~\textdom{val}\rightarrow \bool \\
  \enoughburnsexpr~(v : val) &= \enoughburnsval~v \\
  \enoughburnsexpr~(\RecE\lvarF(\lvar)= \expr) &= (\enoughburnsexpr~\expr = 0)~\wedge \\
                                               &\qquad \enoughburnsexpr~\expr \\
  \enoughburnsexpr~(x : var) &= true \\
  \enoughburnsexpr~\NewProph &= true \\
  \enoughburnsexpr~e &= \bigwedge_{i=1}^{n} \enoughburnsexpr~\expr_i \\
    & \qquad \text{where $\expr_1 \ldots \expr_n$ are the sub-expressions of $e$} \\
  \\
  \enoughburnsval~l &= true, \text{where $l$ is a literal value} \\
  \enoughburnsval~(\RecV\lvarF(\lvar)= \expr) &= (\enoughburnsexpr~\expr = 0)~\wedge \\
                                              &\qquad \enoughburnsexpr~\expr \\
  \enoughburnsval~(\val,\valB)_\valForm &= \enoughburnsval~\val \wedge \\
                                        &\qquad \enoughburnsval~\valB \\
  \enoughburnsval~(\Inl(\val)_\valForm) &= \enoughburnsval~\val \\
  \enoughburnsval~(\Inr(\val)_\valForm) &= \enoughburnsval~\val \\
\end{align*}
  \caption{Determining if there are enough burns in an expressions/value.}%
  \label{fig:enough-burns-def}
\end{figure}

\begin{figure}
  \begin{align*}
    \enoughburnsheap~&:~\textdom{State} \rightarrow \textdom{Prop} \\
    \enoughburnsheap~\state &= \forall l, v.\; \state.\stateHeap(l) = v \rightarrow (\enoughburnsval~v) = true \\
  \end{align*}
  \caption{Ensure that all functions in the heap have enough burns (see Figure~\ref{fig:enough-burns-def}).
  $\textdom{Prop}$ refers to the \coq type.}%
  \label{fig:enough-burns-heap-def}
\end{figure}

\begin{figure}
\begin{align*}
  \forall \rho.\;
  (\forall e \in \rho.\textsc{threads}.~\enoughburnsexpr e = true) \wedge \enoughburnscfg~\rho.\textsc{state}.\stateHeap
\end{align*}
\caption{Requirement on program configurations to have enough burns. Each thread's expression in the configuration $\rho$ needs to have enough\_burns. Moreover all expression on the heap need to have enough burns as well.}%
\label{fig:enough-burns-cfg-def}
\end{figure}

In this section we describe our approach to guarantee termination of programs.

Intuitively a program terminates if it \emph{eventually} reduces to a value, or gets stuck.
Equivalently, we can prove the impossibility of an infinite execution which we define as follows:
\begin{defn}[Infinite execution]
  We say a configuration $\rho$ diverges, written $\mathsf{div}\ \rho$, if there exists an infinite sequence of configurations, starting with $\rho$, where each configuration is related to the next by the small-step relation.
\[ \mathsf{div}~\rho \eqdef \exists (\tau : \nat \rightarrow \textdom{cfg}).\;\forall n. ({\tau~n} \tpstep[\vec\obs] {\tau~(n + 1)}) \wedge (\tau~0 = \rho) \]
\end{defn}
We show that having inserted ``enough'' \burn instructions (see Figure~\ref{fig:enough-burns-cfg-def}, Section~\ref{sec:termination-proof-sketch}), programs are guaranteed to terminate.
Currently, we overapproximate the property of a program having enough \burns by essentially requiring every function application to be syntactically preceded by a \burn.
See Section~\ref{sec:termination-proof-sketch} for more details.
If we can prove in Iris that the instrumented program containing enough \burns does not get stuck, we know that erasing the \burn instructions yields a program reducing to the same value\footnote{This part has not been proven in the \coq development yet.}.

The proof of termination follows the classic approach of assigning to every program state a measure (``size'') equipped with a well-founded relation.
By showing that every step of the program reduces the well-founded relation, it becomes clear that infinite executions are impossible (cf.~\autocite{DBLP:journals/annals/MorrisJ84} and~\autocite{floyd1993}).

\subsection{Proof sketch}%
\label{sec:termination-proof-sketch}

All \heaplanglt operations except for function application are unproblematic with regard to termination.
Consider the measure defined in Figure~\ref{fig:pseudo-size-def}.
We prove the following lemma that illustrates this point.
\begin{lem}[Pseudo size decreases]
  For all expressions except function application, a successful head-step from $\expr$ to $\expr'$ will reduce the pseudo size: $\pseudosize~\expr > \pseudosize~\expr'$.
\end{lem}
The proof is straightforward, proceeding by structural induction on expressions.
It can be found in the code accompanying this report.

To handle function application, we make use of \heaplanglt's call permissions and its \burn construct (Section~\ref{sec:heaplanglt-call-permissions}).
Our syntactic check(s) defined in Figures~\ref{fig:enough-burns-def}, \ref{fig:enough-burns-heap-def} and~\ref{fig:enough-burns-cfg-def} guarantee that there does not exist a function that includes ``unprotected'' function applications.
A function application is unprotected if it is reachable in the control flow graph of the program without passing through a (wrapping) \burn instruction first.
The function given in Figure~\ref{fig:nb-unprotected-apps-def} and its use in Figure~\ref{fig:enough-burns-def} overapproximates this number.

A program thus instrumented---i.e.\ having enough burns as determined by the check of Figure~\ref{fig:enough-burns-cfg-def}---reduces the well-founded relation shown in Figure~\ref{fig:size-definition} at every step of the program, thus precluding infinite executions.
The triple of 1) call permissions, 2) number of unprotected applications and 3) pseudo size is ordered lexicographically (Figure~\ref{fig:size-definition}):
\begin{itemize}
  \item When reducing the stock of call permissions (a non-stuck \burn) we are allowed to arbitrarily (but finitely) increase the number of \emph{now allowed} unprotected function applications.
  \item By decreasing the natural number indicating the allowed number of unprotected function applications, we are allowed to arbitrarily (but finitely) increase the pseudo size, thus justifying the increase in pseudo size incurred by the substitution of function application.
  \item Finally, all other expressions simply reduce the pseudo size.
\end{itemize}

\begin{figure}
\centering
\begin{tabular}{ccccc}
  $\textdom{Multiset}$ & $\times$ & $\nat$ & $\times$ & $\nat$ \\
  $\mathit{mult}$ && $<$ && $<$
\end{tabular}
\caption{The program size is measured using this triple.
The respective well-founded orders shown on the second line are used to define the likewise well-founded lexicographic order (significance decreasing from left to right) over this triple: $<_{\mathit{size}}$. $\mathit{mult}$ refers to the multiset order defined in Figure~\ref{fig:mult1-original}}%
\label{fig:size-definition}
\end{figure}

The well-founded multiset order we use in the proofs (``$\mathit{mult}$'', Figure~\ref{fig:mult1-original}) is a translation of \citeauthor{DBLP:conf/rta/BlanchetteFT17}'s definition (Figure~\ref{fig:mult1-original}) to \coq.
While the well-foundedness of this multiset order has been proven in both Isabelle/HOL~\autocite{isabelle-theory-multiset} and \coq~\autocite{coq-color-multiset-order} it is currently admitted in our development.

\begin{figure}
  \begin{align*}
    \mathit{mult1}~R = \{ (A,B).\; \exists y, B_0, X.\; B = B_0 \uplus [y] \wedge A = B_0 \uplus X \wedge \forall x \in X.\; (x,y) \in R \}
  \end{align*}
  \caption{The one-step Dershowitz-Manna extension $\mathit{mult1}$~\autocite{DBLP:conf/rta/BlanchetteFT17}. $\mathit{mult}$ is then defined as the transitive closure of $\mathit{mult1}$. The syntax has been slightly adapted.}%
  \label{fig:mult1-original}
\end{figure}

Next we will discuss a few key theorems and lemmas, proving the soundness of our syntactic check i.e.\ that $\enoughburnscfg$ prevents infinite executions.
\begin{thm}[No infinite executions]
\[ \forall \rho.\; \enoughburnscfg~\rho \implies \neg (\mathsf{div}~\rho) \]
\end{thm}
This theorem proves the soundness of our approach.
Its proof follows directly from the well-foundedness of the order shown in Figure~\ref{fig:size-definition} and the following lemma.
\begin{lem}[Program steps decrease the program size and preserve enough burns]
  If the starting state has enough \burns, then every possible step decreases the program size.
  Moreover, the next state will still satisfy our syntactic check.
\begin{align*}
  \forall \rho. \enoughburnscfg~\rho\implies \rho \tpstep[\vec\obs] \rho' \implies \\
  \mathit{size}~\rho' <_\mathit{size} \mathit{size}~\rho \wedge \enoughburnscfg~\rho'
\end{align*}
\end{lem}
Its proof is built up from many smaller lemmas following the construction of \heaplanglt.
The program-level step relation is defined using primitive steps, which are defined using head steps and evaluation contexts.
For each of these language-definition levels (no relation to the call permission levels) we essentially prove the same theorem, afterwards lifting it to the next higher level.
Moreover, multiple lemmas have to be shown to prove that substitution and filling evaluation contexts preserves the constituent parts of our definition of $\enoughburnsexpr$.

\chapter{Program logic}%
\label{sec:heaplanglt-program-logic}

We instantiate the Iris framework with our modified version of \heaplang: \heaplanglt.
Given the operational semantics of our language, we reuse Iris' parameterized definitions of weakest preconditions and its adequacy result\footnote{With minor adaption to account for call permissions.}~\autocite{iris-ground-up}.
Moreover, because we only add to \heaplang, we inherit the existing reasoning principles, tactics, notation etc.\ of \heaplang with minor to no adaptations.

Iris' weakest precondition uses a ``state interpretation'', that is an invariant about the state, to be parametric in the way that the language's physical resources are reflected in the logic.
To wit, the language's ``physical'' state is forced to agree with a step-indexed, logical representation, thus giving us the ability to reason about the physical state with the very flexible notion of ``ghost resources'' employed by Iris.
In order to reason about the stock (multiset) of call permissions added to \heaplanglt's state (Figure~\ref{fig:state-def}), they must be reflected in the state interpretation in a suitable manner, as shown next.

\section{Call permissions}

Conveniently, Iris already ships with the definition for a multiset union (unital) ``camera''\footnote{``[A] camera can be thought of as
a kind of `step-indexed PCM [Partial Commutative Monoid]'\,''~\autocite{iris-ground-up}}.
It uses \( \uplus \) (disjoint union) as the camera's operation, the empty multiset \( \empty \) as core (cf.\ \autocite{iris-ground-up}), and \( \subseteq \) as the camera-inclusion operation.

We can simply wrap this multiset union camera ($\textdom{Multiset}$(\textdom{T}), for any countable element type~$\textdom{T}$) with the $\authm$ camera, thus obtaining the ability to talk about \emph{the} authoritative multiset ($\authfull M$) (at some ghost location $\gname$) and its fragment ($\authfrag M$) (also at ghost location $\gname$).
Because we use \( \uplus \) as the underlying operation of the multiset camera, the multiset fragment(s) can be split and combined accordingly.
As a trivial example, we can turn ownership of $\authfrag~[a,a] = \authfrag~([a] \uplus ~[a])$, into ownership of the two fragments $\authfrag~[a]~*~\authfrag~[a]$.
Our proof of termination for a concurrent stack with helping (Section~\ref{sec:concurrent-stack-with-helping}) relies crucially on this property of the $\authm(\textdom{Multiset})$ camera.

From this general composite camera construction we create a singleton authoritative multiset of call permissions.
Its authoritative part is managed by the weakest precondition of Iris and tied to the state of the program via the state interpretation.
The singleton's fragments can be manipulated as expected at the program logic level.
This setup mirrors the way the heap and reasoning about it is handled in standard Iris~\autocite{iris-ground-up}).
In principle the authoritative multiset camera allows us to add arbitrary elements to it if we control the authoritative part.
However, the state interpretation's ownership of the authoritative part, and its use in the definition of the weakest precondition (cf.~\autocite[p.~60]{iris-ground-up}) forces us to match up \heaplanglt's ``physical'' multiset of call permissions and its Iris-level logical representation at every step, thus preventing abuse.

\subsection{Weakest precondition of \langkw{Burn}}%
\label{sec:wp-burn}

The following lemma abstracts away the explicit reasoning about the weakest precondition based on its definition.
\begin{lem}[Weakest precondition of \burn]
\[ 0 \leq n \rightarrow
cp' < cp \rightarrow
\authfrag [cp]\;*\;
(\later (\authfrag [n \cdot cp'] \wand wp~e~\{\Phi\})) \wand
wp~(\langkw{Burn}~\expr~cp~n~cp') \{\Phi\} \]
\end{lem}
Its proof relies on the weakest precondition's access to the state interpretation and its authoritative part ($\authfull$) for the stock of call permissions.
Together with the fragment ($\authfrag$) of the premise, it is able to remove the burned call permission and insert the newly created call permissions.
The premises match the side conditions of \langkw{Burn} (cf.\ Figure~\ref{fig:heaplang-reduction-impure}) with the authoritative fragment guaranteeing that the thread executing the \langkw{Burn} has exclusive access to the call permission it is allowed to burn.
Thus, the \langkw{Burn} instruction is guaranteed to not get stuck.
Lastly, if the expression wrapped by the \langkw{Burn} satisfies the postcondition having only access to the lowered call permissions, the entire expression will satisfy its postcondition if the program terminates (proven separately).

\section{Atomic blocks}

For the time being, weakest preconditions of \langkw{AtomicBlock}s are proven immediately from the $\mathit{lifting.wp\_lift\_atomic\_step}$ lemma of Iris, thus more or less directly working with the definition of weakest precondition.
A supporting lemma similar to that of \langkw{Burn} (Section~\ref{sec:wp-burn}) is left as future work.

\chapter{Example: A concurrent stack with helping}%
\label{sec:concurrent-stack-with-helping}

We validate our approach by using it to modularly (cf.\ Figure~\ref{fig:example-spec}) verify the termination of the operations on a concurrent stack with helping.

The concurrent stack with helping we consider~\autocite{iris-concurrent-stacks-with-helping,iris-concurrent-stacks-with-helping-source-code} is an example of a fine-grained, lock-free, concurrent data structure.
Instead of requiring exclusive access to the entire stack for the complete duration of each operation---thus synchronizing accesses in a coarse-grained manner and ensuring the safe operation of the stack---the atomic instruction ``compare-and-swap'' (CAS)%
\footnote{In the modified code that we prove, we have replaced the primitive CAS with an atomic block of the same effect to show that our \langkw{AtomicBlock} construct can be used to model common atomic hardware primitives.}%
is used for synchronization.
The concurrent stack is lock-free (a particular kind of non-blocking) in the sense that ``it ensures [\ldots] that some thread always makes progress''~\autocite{DBLP:conf/icdcs/HerlihyLM03}.
By contrast, in a blocking algorithm threads may rely on interference by other thread(s) to make progress.

\begin{figure}
\center
\includegraphics[width=.7\textwidth]{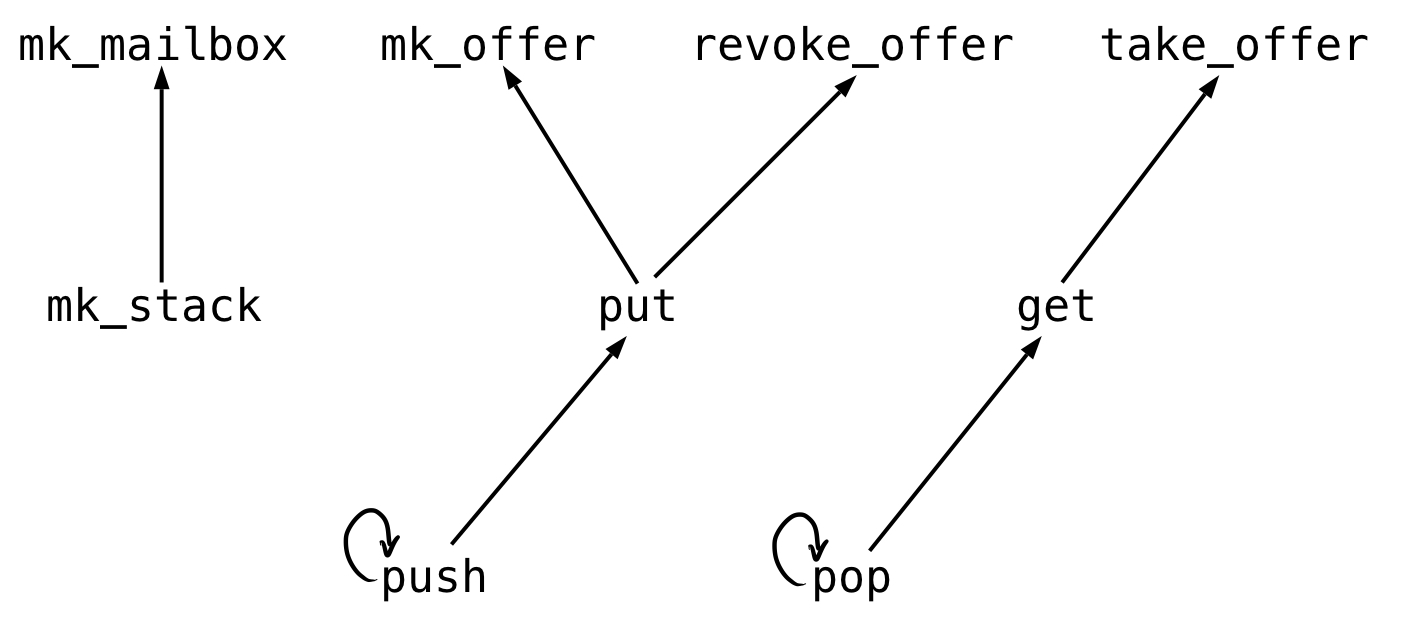}
\caption{Static call graph of concurrent stack with helping module.}%
\label{fig:example-static-call-graph}
\end{figure}

The concurrent stack in question uses ``helping'' (inter-thread cooperation) internally to reduce contention:
in addition to the actual stack, implemented as a linked list, a side-channel (``mailbox'') is managed so that a pushing and popping thread can exchange values directly without involving the underlying linked list.
Proving this internal (not exposed in the specification shown in Figure~\ref{fig:example-spec}) optimization is non-trivial~\autocite{iris-concurrent-stacks-with-helping}.
Importantly, helping is present in a ``significant class of sophisticated concurrent data structures'' but is not supported by some other major program logics~\autocite[p.\ 2]{iris}.
For our \emph{adaptation}, however, the helping employed did not pose a significant challenge to verification of termination.
Specifically, the parts of the original proof (\autocite{iris-concurrent-stacks-with-helping,iris-concurrent-stacks-with-helping-source-code}) related to handling the helping could essentially be reused verbatim.
To see why termination verification for these parts is not an issue, consider the static call graph of our concurrent stack shown in Figure~\ref{fig:example-static-call-graph}.
All functions, except for \texttt{push} and \texttt{pop}, only perform ``upcalls''; it is therefore easy to verify that these functions terminate with our level-based argument.
Consequently, we will focus on the proof of \texttt{push} and refer the reader to the \coq development for the proofs of the remaining functions%
\footnote{With the exception of \texttt{pop}, which is admitted. However, we are confident that the reasoning techniques employed for the verification of \texttt{push} scale to \texttt{pop}.}
.

\begin{figure}
\begin{align*}
& \{ P~[~~]~\textcolor{blue}{*~\authfrag [\mathsf{STACK\_OP}]} \}~\texttt{new\_stack \#()}~\{\texttt{v}, RET~\texttt{v}; is\_stack~P~\texttt{v} \} \\
\\
& \{ is\_stack~P~\texttt{s}~*(\forall xs.\; P~xs \vsW[][\top\setminus N\uparrow] P (\texttt{v} :: xs) * \Psi  \texttt{\#()})~\textcolor{blue}{*~\authfrag [\mathsf{STACK\_OP}]} \} \\
& \texttt{stack\_push~s~v} \\ % STACK PUSH
& \{ RET~\texttt{\#()}, \Psi \texttt{\#()} \} \\
\\
& \{ is\_stack~P~s~*
((\forall \texttt{v}, xs.\; P~(\texttt{v} :: xs) \vsW[][\top\setminus N\uparrow] P~xs * \Psi~(\textsc{somev}~\texttt{v})) \wedge
(P~[~~] \vsW[][\top\setminus N\uparrow] P~[~~]~* \\
& \qquad \Psi~\textsc{nonev}))~\textcolor{blue}{*~\authfrag [\mathsf{STACK\_OP}]} \} \\
& \texttt{stack\_pop~s} \\ % STACK POP
& \{ \texttt{v}, RET~\texttt{v}; \Psi~\texttt{v} \}
\end{align*}
\caption{Predicate-based HoCAP~\autocite{hocap} style spec taken from the Iris examples repo~\autocite{iris-concurrent-stacks-with-helping-source-code}.
Our additions required to prove termination are highlighted in blue.
In this case the HoCAP style spec proven here and a proof of logical atomicity are equivalent~\autocite{iris-hocap-pred-hocap-auth-logatom-equiv-for-concurrent-stack}.
$N$ refers to the namespace of the stack invariant (see Figure~\ref{fig:stack-invariant}). Its removal from the full ``mask'' ($\top$) is necessary to open the invariant. Masks and namespaces are Iris' way of preventing reentrancy; see \autocite[p.\ 14]{iris-ground-up} for more details.
}%
\label{fig:example-spec}
\end{figure}

\begin{figure}
\begin{align*}
& stack\_inv~P~l~\textcolor{blue}{l'~\gamma} \eqdef \exists v, xs \textcolor{blue}{, n, \mathit{in\_progress}}.\; l \mapsto v~*~\mathit{is\_list}~xs~v~*~P~xs~\textcolor{blue}{*}\\
& \qquad \textcolor{blue}{\ownGhost{\gamma}{\authfull~\mathit{in\_progress}} * l' \mapsto n * size(\mathit{in\_progress}) \leq n~*} \\
& \qquad \textcolor{blue}{\authfrag~[\mathit{size}([x \in \mathit{in\_progress }~|~x \neq v]) \cdot \mathsf{CP\_TRY}]} \\
& \\
& \mathit{is\_stack}~P~v~\eqdef \exists \mathit{mailbox}, l, \textcolor{blue}{l', \gamma}.\; v = (\mathit{mailbox}, l, \textcolor{blue}{l'})~*~\mathit{is\_mailbox}~P~\mathit{mailbox}~* \\
& \qquad \knowInv{N}{\mathit{stack\_inv}~P~l~\textcolor{blue}{l'~\gamma}}
\end{align*}
\caption{Modified stack invariant. Changes are highlighted in blue.
Note that now there is another pointer, which stores an upper bound on the number of concurrent threads accessing the stack at any given time.
The $\mathit{in\_progress}$ multiset stores each accessing thread's view of the stack's head at the beginning of their current stack operation.
For all threads that have an incorrect view (because of an invalidating operation by another thread) a compensating call permission is stored.
}%
\label{fig:stack-invariant}
\end{figure}

While we do have to change the code (Figure~\ref{fig:push-inner-code})---to add our \burn instructions---and modify the stack invariant (Figure~\ref{fig:stack-invariant}), these changes are hidden behind the specification of Figure~\ref{fig:example-spec} with the exception of the added call permissions shown in blue.

We refer the reader to the original case study~\autocite{iris-concurrent-stacks-with-helping} and its current online sources~\autocite{iris-concurrent-stacks-with-helping-source-code} for the proof of (partial) functional correctness.
Here, we will focus on our changes to the code and proofs to enforce safe termination.
The code uses the following notation for a \( \langkw{Burn}~\expr_1~cp~\expr_2~cp' \) expression:
\texttt{burn} $cp$ \texttt{receive} $\expr_2$ \texttt{times} $cp'$ \texttt{in} $\expr_1$.
Additionally, when the multiset of call permissions to receive is empty we omit it: \texttt{burn} $cp$ \texttt{in} $\expr_1$.

\paragraph{High-level proof overview}
Proving the termination of the stack's push operation is non-trivial because when it fails due to interference from another thread, it will \emph{retry} by recursively calling itself again with the same arguments.
Simply put, there is no straightforward termination measure:
\begin{itemize}
  \item the syntactic size (and $\pseudosize$) of the expression increases
  \item the function application is not an ``upcall''
  \item the function's arguments do not structurally decrease\footnote{We do currently not directly support such an argument (see Section~\ref{sec:overapprox-finite-acyclic-ds}).}
\end{itemize}
Instead, we exploit the property that when \texttt{push} is interfered with, another thread must have made progress i.e.\ successfully modified the stack.
By ensuring that the (successfully) interfering thread decreases the global termination measure (cf.\ Section~\ref{sec:heaplanglt-termination}) ``appropriately'' (details below), we can compensate all interfered-with threads.
Consequently, the current thread will either be fortunately scheduled and progress towards termination early---potentially compensating other threads along the way---or be the last thread running at which point no other thread can interfere.
This argument critically relies on our syntactic check that prevents infinite executions in general and therefore unbounded interference as a corollary.
It is reminiscent of \citeauthor{DBLP:conf/esop/PintoDGS16}'s ordinal bound assigned to concurrent stacks, which describes the amount of interference that a stack can sustain~\autocite{DBLP:conf/esop/PintoDGS16}.

\begin{figure}
\includegraphics[width=\linewidth]{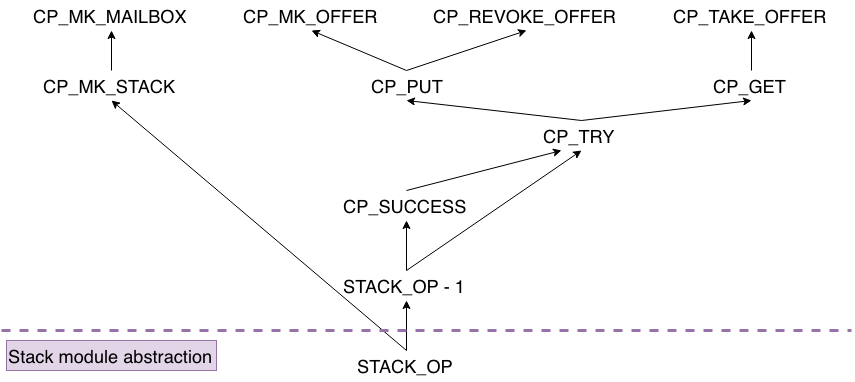}
\caption{Well-founded order of call permissions used in our development.
There is a strong connection to the static call graph in Figure~\ref{fig:example-static-call-graph}.
The recursive call of \texttt{push} and \texttt{pop} are encoded with $\mathsf{CP\_SUCCESS}$ and $\mathsf{CP\_TRY}$ as described in Section~\ref{sec:concurrent-stack-with-helping}.
The \coq formalization currently uses appropriately picked natural numbers.}%
\label{fig:call-permission-ordering}
\end{figure}

Our compensation scheme is reflected in the modified stack invariant shown in Figure~\ref{fig:stack-invariant}.
In addition to guaranteeing the functional correctness of the stack, our extended stack invariant also ensures the following two additional properties of our compensation scheme:
\begin{enumerate}
  \item Successfully interfering threads need to ``compensate'' all possibly-interfered-with threads.
  \item Interfered-with threads must be able to prove that they have been compensated in order to retry.
\end{enumerate}
For both of these properties the stack invariant makes use of a per-stack authoritative multiset ($\mathit{in\_progress}$) that keeps track of the stack operations in progress.
In order to be able to retry its operation upon being interfered with, a thread needs to first register itself in this multiset.
Because a thread does not know a priori whether it will be interfered with, each thread performing a stack operation needs to register itself.
Registering consists of inserting a thread's current view of the stack's linked list's head pointer into the multiset $\mathit{in\_progress}$.
Upon registering, a thread receives a ``receipt'' (the fragment for the inserted element).
It is impossible to remove that particular element without the matching receipt therefore guaranteeing that the thread would be accounted for upon interference.
The compensation itself takes the form of call permissions at level $\mathsf{CP\_TRY}$.
The set-builder-notation-like expression $[x \in \mathit{in\_progress }~|~x \neq v]$ filters out the registrations that have been invalidated by a successfully interfering thread i.e.\ all the threads whose ``view'' on the concurrent stack is now wrong and which need to retry.
For each of those invalidated registrations a call permission of level $\mathsf{CP\_TRY}$ must be deposited in the stack invariant by the successful thread: $\authfrag~[\mathit{size}([x \in \mathit{in\_progress }~|~x \neq v])~\cdot \mathsf{CP\_TRY}]$.
So far the setup is identical to that of \citeauthor{DBLP:journals/toplas/JacobsBK18}~\autocite{DBLP:journals/toplas/JacobsBK18}.
However, unlike them we cannot choose the amount of created call permissions directly in the program logic (see Section~\ref{sec:overview}) so we introduce an additional memory location that stores an overapproximation on the size of the $\mathit{in\_progress}$ multiset ($l'$ in the stack invariant shown in Figure~\ref{fig:stack-invariant}, \texttt{ln} in the code shown in Figure~\ref{fig:push-inner-code}).
After successfully modifying the stack, the number stored at that location can then be used to create enough call permissions to compensate all interfered-with threads (Line~\ref{line:compensate-threads} in Figure~\ref{fig:push-inner-code}).
For this to work, registration in the $\mathit{in\_progress}$ multiset must be accompanied by an increment of that counter (Line~\ref{line:increase-upper-bound} in Figure~\ref{fig:push-inner-code}).
We note that from the point of the view of the stack's specification (Figure~\ref{fig:example-spec}) and therefore the stack's clients, all these details are hidden behind the call permissions at level $\mathsf{STACK\_OP}$ associated with the stack module (cf.\ Figure~\ref{fig:call-permission-ordering}).

Below, we summarize the changes to interactions with the stack invariant due to the encoded compensation scheme.
\begin{itemize}
    \item The helping code is not impacted because the relevant $is\_mailbox$ predicate and the invariant contained within (not shown) are unmodified.
    \item \emph{Stack creation}: No thread can interfere, therefore no call permissions need to be deposited yet. We can simply allocate an empty $\mathit{in\_progress}$ multiset before establishing the modified stack invariant.
    \item \emph{Stack modification}:
      Whenever a stack operation wants to modify the underlying linked list, it adds itself to the $in\_progress$ multiset with its current view of the linked list's head.
      It receives a corresponding fragment, establishing that it is the only thread that can remove that specific ``registration''.
      These are program-logic-level operations and not visible in the code itself.
      Their approximate location has been highlighted with comments in Figure~\ref{fig:push-inner-code}.
      \begin{itemize}
          \item Failure: (if-statement beginning at Line~\ref{line:if-cas} returns \texttt{\#false})
            The registered view on the linked list's head was invalidated.
            Consequently, we know that there is at least one call permission in the stack invariant, as evidenced by the registration fragment.
            By removing the registration from the $in\_progress$ multiset, a corresponding call permission can be removed and the stack invariant reestablished.
            The removed call permission is required in order to retry the operation (Figure~\ref{fig:push-inner-code}, Line~\ref{line:retry}).
          \item Success: (if-statement beginning at Line~\ref{line:if-cas} returns \texttt{\#true})
              Upon success, the higher call permission ($\mathsf{CP\_SUCCESS}$) is burned to receive $n$ call permissions at level $\mathsf{CP\_TRY}$ where $n$ is the upper bound on the number of possibly-interfered-with threads as tracked by the stack invariant.
              This allows us to change the linked list's head and re-establish the stack invariant.
      \end{itemize}
\end{itemize}

This proof sketch is reflected in our changes to the code of \texttt{push}, renamed to \texttt{push\_inner}, shown in Figure~\ref{fig:push-inner-code}
We briefly relate the changed code to the proof sketched above.
Figure~\ref{fig:push-code} wraps \texttt{push\_inner} in order to satisfy the modular specification of Figure~\ref{fig:example-spec}.
\begin{itemize}
  \item Lines~\ref{line:begin-unpack}--\ref{line:end-unpack}:
    The stack is now a triple of mailbox, linked list location and the location tracking the upper bound on the number of possibly-interfered with threads.
    This reflects the additional location tracked by the stack invariant (Figure~\ref{fig:stack-invariant}).
  \item Line~\ref{line:burn-cp-try}: Burn $\mathsf{CP\_TRY}$ to receive two call permissions of level $\mathsf{CP\_PUT}$ (used on lines~\ref{line:burn-cp-put-1} and~\ref{line:burn-cp-put-2}).
  \item Line~\ref{line:burn-cp-put-1}: Burn call permission to justify call to \texttt{put}.
  \item Line~\ref{line:increase-upper-bound}: Increase upper bound on the number of possibly-interfered threads.
    Additionally, at the ghost resource level (at the program logic level), the current head of the linked list (\texttt{!"s"}) is added to $in\_progress$, thus receiving the registration fragment.
    The stack invariant is updated to reflect these changes.
    To do so, we use our \langkw{AtomicBlock} construct.
    Note that after erasing the \burn, the atomic block becomes obsolete because the load operation is defined to only take one atomic step in \heaplanglt.
  \item Lines~\ref{line:manual-cas-start}--\ref{line:manual-cas-end}: Manual implementation of the original CAS.
    This demonstrates that CAS can be naturally encoded with our atomic block construct.
  \item Lines~\ref{line:cas-will-succeed-start}--\ref{line:cas-will-succeed-end}: The ``CAS'' was successful; the $\mathsf{CP\_SUCCESS}$ call permission is turned into \texttt{!"ln"} call permissions at level $\mathsf{CP\_TRY}$, justifying the interfered-with threads' retry operations.
    Reading the upper bound and updating the stock of call permissions needs to happen atomically with the CAS (or our manual translation) so as not to miss any threads.
  \item Line~\ref{line:burn-cp-put-2}: Burn call permission at level $\mathsf{CP\_PUT}$ (obtained on Line~\ref{line:burn-cp-try}) to justify the retry operation.
    However, if it was not for the $\mathsf{CP\_TRY}$ obtained on Line~\ref{line:receive-cp} as compensation we would not be able to finish the proof.
    In order to complete the proof we need to apply Löb induction which requires us to show that we can satisfy \texttt{push\_inner}'s precondition.
    Due to the compensation scheme, we were able to remove a $\mathsf{CP\_TRY}$ call permissions from the stack invariant by removing the registration from the multiset $in\_progress$ (Line~\ref{line:receive-cp}), justified by turning in the receipt ($\authfrag [\texttt{tail}]$) obtained after registering ($\authfrag [t]$, cf.\ description for Line~\ref{line:increase-upper-bound} above).
    The postcondition is satisfied (Line~\ref{line:post-loeb}) after application of Löb induction (see~\autocite{iris-ground-up}).
  \item Figure~\ref{fig:push-code} Wrapper function to hide the detailed termination argument just described.
\end{itemize}

\begin{figure}
\begin{lstlisting}[escapechar=@]
@\( \textcolor{brown}{ \{ is\_stack~P~p~*(\forall xs.\; P~xs~\vsW[][\top\setminus N\uparrow] P (v :: xs) * \Psi~\texttt{\#()})*\authfrag [\mathsf{CP\_TRY}; \mathsf{CP\_SUCCESS}] \} } \)@ @\label{line:push-inner-precondition}@
rec: "push_inner" "p" "v" :=
  let: "mailbox" := Fst @\textcolor{blue}{\$ Fst}@ "p" in @\label{line:begin-unpack}@
  let: "s" := Snd @\textcolor{blue}{\$ Fst}@ "p" in
  @\textcolor{blue}{let: "ln" := Snd "p" in}@ @\label{line:end-unpack}@
  @\textcolor{blue}{burn $\mathsf{CP\_TRY}$ receive \#2 times $\mathsf{CP\_PUT}$ in}@ @\label{line:burn-cp-try}@
  (match: @\textcolor{blue}{burn $\mathsf{CP\_PUT}$ in}@ put "mailbox" "v" with @\label{line:burn-cp-put-1}@
    NONE => #()
  | SOME "v'" =>
    let: "tail" := @\textcolor{gray}{// Open stack invariant}@
      @\textcolor{blue}{AtomicBlock (}@
      @\textcolor{brown}{$\{\texttt{s} \mapsto \mathit{t} * \texttt{ln} \mapsto n * \mathit{size}(\mathit{in\_progress}) \leq n * \ownGhost{\gamma}{\authfull~\mathit{in\_progress}}\}$}@
      @\textcolor{blue}{"ln" <- !"ln" + \#1;; }@ @\label{line:increase-upper-bound}@
      @\textcolor{brown}{$\{\texttt{s} \mapsto t * \texttt{ln} \mapsto n + 1 * \mathit{size}(\mathit{in\_progress} \uplus [t]) \leq n + 1 * \ownGhost{\gamma}{\authfull (\mathit{in\_progress} \uplus [t])} * \ownGhost{\gamma}{\authfrag [t]}\}$}@
      !"s" @\textcolor{blue}{)}@ @\textcolor{gray}{// Close stack invariant}@
    in
    let: "new" := SOME (ref ("v'", "tail")) in
    if: @\textcolor{gray}{// Open stack invariant}@ @\label{line:if-cas}@
        @\textcolor{red}{CAS("s" "tail" "new")}@ @\textcolor{blue}{AtomicBlock (} \textcolor{gray}{// manual CAS}@ @\label{line:manual-cas-start}@
          @\textcolor{blue}{let: "vl" := !"s" in}@
          @\textcolor{blue}{if: "vl" = "tail" then (} \textcolor{gray}{// Success!}@
            @\textcolor{brown}{$\{ \authfrag [\mathsf{CP\_SUCCESS}] * \texttt{ln} \mapsto n * \mathit{size}(\mathit{in\_progress}) \leq n \}$}@ @\label{line:cas-will-succeed-start}@
            @\textcolor{blue}{burn $\mathsf{CP\_SUCCESS}$ receive \texttt{!"ln"} times $\mathsf{CP\_TRY}$ in}@ @\label{line:compensate-threads}@
            @\textcolor{brown}{$\{ \authfrag [n \cdot \mathsf{CP\_TRY}] * \texttt{ln} \mapsto n * \mathit{size}(\mathit{in\_progress}) \leq n \}$}@
            @\textcolor{blue}{("s" <- "new";; \#true)}@ @\label{line:cas-will-succeed-end}@
          @\textcolor{blue}{) else}@
            @\textcolor{brown}{$\{ \ownGhost{\gamma}{\authfull \mathit{in\_progress}} * \ownGhost{\gamma}{\authfrag [\texttt{tail}]} * \}$}@
             @\textcolor{brown}{$ \authfrag [\mathit{size}([x \in \mathit{in\_progress } | x \neq \texttt{vl}]) \cdot \mathsf{CP\_TRY}] \}$}@
            @\textcolor{brown}{$\{ \authfrag [\mathsf{CP\_TRY}] * \ownGhost{\gamma}{\authfull (\mathit{in\_progress} \setminus [\texttt{tail}])} * $}@ @\label{line:receive-cp}@
             @\textcolor{brown}{$ \authfrag [\mathit{size}([x \in (\mathit{in\_progress} \setminus [\texttt{tail}]) | x \neq \texttt{vl}]) \cdot \mathsf{CP\_TRY}] \}$}@
            @\textcolor{blue}{\#false}@
        @\textcolor{blue}{)}@ @\textcolor{gray}{// Close stack invariant}@ @\label{line:manual-cas-end}@
    then #()
    else
      @\textcolor{blue}{burn $\mathsf{CP\_PUT}$ in}@ @\label{line:burn-cp-put-2}@
      @\( \textcolor{brown}{ \{ is\_stack~P~p*(\forall xs.\; P~xs~\vsW[][\top\setminus N\uparrow] P (v :: xs) * \Psi \texttt{\#()})*\authfrag [\mathsf{CP\_TRY}; \mathsf{CP\_SUCCESS}] \} } \)@
      "push_inner" "p" "v'" @\textcolor{gray}{// Apply Löb induction}@ @\label{line:retry}@
      @\( \textcolor{brown}{ \{ RET~\texttt{\#()}, \Psi~\texttt{\#()} \} } \)@ @\label{line:post-loeb}@
  end).
@\( \textcolor{brown}{ \{ RET~\texttt{\#()}, \Psi~\texttt{\#()} \} } \)@
\end{lstlisting}
\caption{Modified code for pushing to the concurrent stack with helping.
Removed code is shown in red, added code is shown in blue, selected proof states in brown.
The code wrapping this function in order to satisfy the specification of Figure~\ref{fig:example-spec} is shown in Figure~\ref{fig:push-inner-code}.
}%
\label{fig:push-inner-code}
\end{figure}

\begin{figure}
\begin{lstlisting}[escapechar=@]
@\( \textcolor{brown}{ \{ is\_stack~P~s~*(\forall xs.\; P~xs~\vsW[][\top\setminus N\uparrow] P (v :: xs) * \Psi  \texttt{\#()})*\authfrag [\mathsf{STACK\_OP}] \} } \)@
@\textcolor{blue}{rec: "push" "p" "v",}@ @\label{line:push}@
  @\textcolor{blue}{burn $\mathsf{STACK\_OP}$ receive \#2 times $\mathsf{STACK\_OP - 1}$ in}@
  @\textcolor{blue}{burn $\mathsf{STACK\_OP - 1}$ receive \#1 times $\mathsf{CP\_TRY}$ in}@
  @\textcolor{blue}{burn $\mathsf{STACK\_OP - 1}$ receive \#1 times $\mathsf{CP\_SUCCESS}$ in}@
  @\( \textcolor{brown}{ \{ is\_stack~P~s~*(\forall xs.\; P~xs~\vsW[][\top\setminus N\uparrow] P (v :: xs) * \Psi  \texttt{\#()})*\authfrag [\mathsf{CP\_TRY}; \mathsf{CP\_SUCCESS}] \} } \)@
  @\textcolor{blue}{push\_inner "s" "v".}@
@\( \textcolor{brown}{ \{ RET~\texttt{\#()}, \Psi~\texttt{\#()} \} } \)@
\end{lstlisting}
\caption{Code wrapping \texttt{push\_inner} of Figure~\ref{fig:push-inner-code}.
  This function only requires $\mathsf{STACK\_OP}$ in its precondition, satisfying the specification of Figure~\ref{fig:example-spec} and thus largely hiding the termination argument ($\mathsf{CP\_SUCCESS}$ and $\mathsf{CP\_TRY}$) which need not concern clients of the stack module.
}%
\label{fig:push-code}
\end{figure}

\chapter{Related work}%
\label{sec:related-work}

The limitations of Iris with regard to liveness properties due to its use of step-indexing are well-known and some workarounds have already been proposed.
\citeauthor{DBLP:conf/esop/TassarottiJ017}~\autocite{DBLP:conf/esop/TassarottiJ017} allows proving termination-preserving refinement when the source language features only bounded non-determinism i.e.\ ``each configuration [cf.\ Section~\ref{sec:heaplanglt-termination}] only has finitely many possible successor configurations.''~\autocite{DBLP:conf/esop/TassarottiJ017}.
Moreover their refinement proofs can only rely on bounded stuttering with a bound that has to be chosen up front~\autocite[p. 2]{DBLP:conf/pldi/SpiesGGTKDB21}.
Termination refinement in this setting ``becomes a safety property, because non-simulation can be determined by examining a finite prefix of execution.''~\autocite{DBLP:conf/pldi/SpiesGGTKDB21}
\citeauthor{DBLP:conf/esop/TassarottiJ017}'s work supports non-blocking and blocking concurrency under fair scheduling.
They have to extend Iris with linear predicates and modify the definition of resource algebras and cameras with a stepping relation~\autocites{DBLP:conf/esop/TassarottiJ017}[p.\ 24]{DBLP:journals/corr/abs-2109-07863}.

\citeauthor{DBLP:journals/corr/abs-2109-07863}~\autocite{DBLP:journals/corr/abs-2109-07863} have a similar finiteness side-condition to \autocite{DBLP:conf/esop/TassarottiJ017}.
However, their notion of history-sensitive refinement is able to prove refinement with respect to an abstract model (state transition system), allowing the user to pick the model and the precise relationship between traces in the model and the program under consideration.
For every step in the program, there must be an extension of the auxillary trace, conforming to the model, corresponding to the program step modulo finite stuttering.
Their extension of Iris, Trillium, is conservative with respect to standard Iris with the exception of prophecies~\autocite{iris:prophecy}, which are not supported.
By appropriately picking the model and proving fair termination for it, they are able to prove termination for programs using blocking concurrency.
Their story with respect to modular specifications and proofs has not been fleshed out yet.

\citeauthor{DBLP:conf/pldi/SpiesGGTKDB21}~\autocite{DBLP:conf/pldi/SpiesGGTKDB21} identify the previously alluded-to existential property as a key property that Iris lacks:
\[ \text{If} \vDash \exists (x : X). \Phi~x\text{, then} \vDash \Phi~x~\text{for some $x : X$} \]
That is, the ability to ``pull out'' the existential qualifier out of the step-indexed logic into the ambient logic.
They propose, and implement, a version of Iris (Transfinite Iris) that uses ordinals instead of natural numbers for the step index.
As motivating example, they prove the termination refinement of a memoizing recursive function with respect to the original non-memoized version.
The employed hash table results in the potential for unbounded stuttering.
Transfinite Iris, however, does not yet support concurrency~\autocite[p. 13]{DBLP:conf/pldi/SpiesGGTKDB21}.
Moreover, two important commuting rules: \textsc{LaterExists} and \textsc{LaterSep} (see Figure~\ref{fig:lost-rules}) are lost in the transfinite model.
It is unclear whether all rules/project using these rules could be re-proven without them.
In particular, the two rules are used for proofs of logical atomicity~\autocite{iris} a major feature of Iris.

\begin{figure}
\begin{mathpar}
  \infer [LaterExists] {\text{$\type$ is inhabited}}
  {\later(\Exists x:\type. \prop) \provesIff \Exists x:\type. \later\prop}

  \infer[LaterSep] {}
  {\later(\prop*\propB) \provesIff \later\prop * \later\propB}
\end{mathpar}
\caption{The two commuting results lost in the transfinite model of Iris.}%
\label{fig:lost-rules}
\end{figure}

\citeauthor{DBLP:conf/esop/MevelJP19}~\autocite{DBLP:conf/esop/MevelJP19}
demonstrate that instrumentation and Iris' resourceful logic allow reasoning about explicit upper bounds on the execution time of programs%
\footnote{In our discussion we will only consider the time credits part of their work, not the time receipts.}%
.
Proving an upper bound on the execution time is of course also a proof of termination.
Like us, they modify the code under consideration: an automatic translation inserts calls to a ``tick'' function that reduces the number held at a fixed location in memory.
That function aborts when the counter reaches zero.
They then prove that the program does not get stuck i.e.\ that it terminates within $n$ steps.

Moving away from Iris to a first-order setting, Total TaDA~\autocite{DBLP:conf/esop/PintoDGS16} extends TaDA~\autocite{DBLP:conf/ecoop/PintoDG14} to support non-blocking concurrency.
TaDA Live~\autocite{DBLP:journals/toplas/DOsualdoSFG21} goes further still and supports termination verification for programs with blocking concurrency under fair scheduling.
With the exception of TaDA---for which a large subset is supported by the mechanized, but not foundational, Voila~\autocite{DBLP:conf/fm/WolfSM21} proof outline checker---there does not exist tool support for these logics.

\citeauthor{DBLP:conf/esop/JaberR21}~\autocite{DBLP:conf/esop/JaberR21} take in some sense the opposite approach to us: instead of extending the programming language under consideration and thus preventing inspection of the step-index this way, they design a refinement type system \emph{on top of} the guarded recursive lambda-calculus.
To the best of our knowledge their work has not been mechanized yet and has not been applied to a concurrent language.

\chapter{Outlook}%
\label{sec:outlook}

\section{Termination of blocking concurrency under fair scheduling}

The presented approach is not able to verify the termination of programs that rely on blocking concurrency under fair scheduling.
An extension of the current development with an adaption of \citeauthor{DBLP:conf/cav/ReinhardJ20}' work~\autocite{DBLP:conf/cav/ReinhardJ20} on ``Ghost Signals'' is planned in order to support blocking concurrency in a modular fashion.
In this approach, threads are allowed to create signals and matching obligations to set those signals.
Signals can be shared freely while obligations may only be transferred during fork operations.
Another level argument ensures that no cyclic dependencies are created; signals and busy-waiting can therefore be used to justify busy waiting under fair scheduling.

\section{Less \burns}%
\label{sec:outlook-less-burns}

Right now, the syntactic check that the program code contains enough \burns is overly restrictive.
Even sequencing and let-bindings, which are syntactic sugar for immediately applied functions, are treated as function applications that must be protected by \burns.
Relaxing this requirement and re-proving the theorem outlined in Section~\ref{sec:heaplanglt-termination} would lessen the annotation burden on the programmer.
More ambitious relaxations, like allowing ``up-calls'' are currently not on the roadmap.
Alternatively, it would be possible to introduce explicit sequencing- and let-constructs to the language instead of defining them as syntactic sugar for function application.

\section{Overapproximation of finite, acyclic, data structures}%
\label{sec:overapprox-finite-acyclic-ds}

An unfortunate shortcoming of the current development is the inability to reason about termination that is justified by traversing finite data structures.
For example, we currently do not support proving that a recursive list length function is terminating.
Intuitively, this should be possible as long as the separation logic predicate describing the finite data structure is strong enough.
Consider the definition of the $is\_list$ predicate from the Iris lecture notes (slightly adapted)~\autocite{iris-lecture-notes}:
\begin{align*}
& \mathit{is\_list}~l~[~] \equiv l = \textsc{NONE} \\
& \mathit{is\_list}~l~(x::\mathit{xs}) \equiv \exists hd, l'.\; l = \textsc{SOME}(\mathit{hd}) * \mathit{hd} \mapsto (x,l') * isList(l',\mathit{xs})
\end{align*}
This predicate's use of (full, not fractional) points-to predicates ensures that the list is acyclic.
Given a program heap defined as finite map, the list cannot be infinite.

An overapproximating approach that allows the \burn construct to inspect the size of the heap seems feasible.
However, exploratory attempts turned out unsuccessful.
While unfortunate, this issue is currently not a priority because it is orthogonal to the challenges of verifying termination of non-blocking and blocking concurrency.

\section{Interaction with other Iris projects}

The current development only relies on standard Iris, instantiated with the slightly adapted example language.
It would be interesting to investigate whether the approach outlined in this report is in practice compatible with existing developments.
Promisingly, the proofs for partial correctness of the concurrent stack in Section~\ref{sec:concurrent-stack-with-helping} were not impacted.
Be simply adapting Iris' default programming language \heaplang, our development inherits Iris' support for prophecies~\autocite{iris:prophecy}.
Nevertheless, we have not yet investigated adapting the proofs for e.g.\ RDCSS~\autocite{DBLP:conf/wdag/HarrisFP02,iris:prophecy} to show termination.
More advanced integrations with modified program logics (e.g.\ \autocite{DBLP:journals/pacmpl/BizjakGKB19}) and different programming languages (e.g. \autocite{DBLP:journals/pacmpl/0002JKD18}) remain to be investigated.

\printbibliography

\end{document}